\documentclass{ws-procs10x7}
\usepackage{epsfig,balance}

\newcolumntype{d}[1]{D{.}{.}{#1}}

\def\Journal#1#2#3#4{{\it #1} {\bf #2}, #3 (#4)}
\newcommand{\met}    {\mbox{${\hbox{$E$\kern-0.6em\lower-.1ex\hbox{/}}}_T$}} %missing ET
\makeindex
\begin{document}

\title{SOME SEARCHES FOR NEW PHYSICS\\
WITH THE D0 DETECTOR}

\author{E. KAJFASZ$^*$ for the D0 Collaboration}

\address{CPPM, CNRS-IN2P3/Universit\'e de la M\'editerran\'ee,\\ Case 902, 13288 Marseille Cedex 9, France\\$^*$E-mail: kajfasz@cppm.in2p3.fr}

%%%%%%%%%%%%%%%%%%%%%%%%%%%%%%%%%%%%%%%%%%%%%%%%%%%%%%%%%%%%%%%%%%%%%%%%%
% You may repeat \author \address as often as necessary                 %
%%%%%%%%%%%%%%%%%%%%%%%%%%%%%%%%%%%%%%%%%%%%%%%%%%%%%%%%%%%%%%%%%%%%%%%%%

\twocolumn[\maketitle\abstract{We present recent results on Technicolor and Leptoquark searches 
obtained analyzing up to 0.4~fb$^{-1}$ of data taken at Fermilab by the D0 experiment during 
the first part of the Tevatron Run II.}
\keywords{Tevatron; D0; Technicolor; Leptoquark.}
]

\section{Introduction}
The Standard Model (SM), although phenomenologically successful, leaves many questions unanswered. To address some 
of these questions, new models and theories have been devised that need to be confronted with experimental facts.
The Tevatron, providing $p\bar p$ collisions at a centre of mass energy of $\sqrt{s}=1.96$~TeV, is 
presently the hadron collider at 
the energy frontier, and thus plays a leading role in the quest of phenomena Beyond the Standard Model (BSM).
Numerous D0 results concerning BSM physics, related to the extension of the Poincar\'e group 
(Supersymmetry and Supergravity)~\cite{susy}, the increase in the number of space dimensions~\cite{xtra}, 
the enlargement the gauge group~\cite{xtra}, and the existence of a substructure to quark and leptons 
(compositeness)~\cite{compos} have 
been shown at this conference. This contribution focuses on searches for
Technicolor, which provides an alternative to the SM Electroweak Symmetry Breaking (EWSB) mechanism, and for 
Leptoquarks, ambivalent particles predicted by several extensions to the SM. A comprehensive list of results of 
D0 search analyses can be found in the experiment web pages\cite{web}.

\section{Search for Technicolor}
%\subsection{Motivations}
In the SM, the Higgs boson field is the key of the spontaneous EWSB mechanism. However, being a scalar particle,
its mass is pushed by radiative corrections towards high energy (GUT or Planck) scales. This gives rise to 
the so-called hierarchy problem which can
be solved {\it e.g.} by taming the quadratically divergent Higgs mass corrections making use of Supersymmetry. 
Alternatively, 
Technicolor (TC) does away with a fundamental scalar, but introduces technifermions subjected to a new stong dynamics 
{\it \`a la} QCD. In the original TC model~\cite{susskind}, the coupling of the unbroken electroweak gauge fields 
to technifermion condensates provides a way
to generate masses only to the $W$ and $Z$ vector bosons~\cite{susskind}. Some extensions~\cite{etc} are necessary to 
make TC more phenomelogically satisfying. 
The analysis presented here is performed in the framework of the Technicolor Strawman Model 
(TCSM2)~\cite{lane}, well suited for the search 
for light technihadrons produced with substantial cross-section at the Tevatron. The lightest technifermions are expected 
to be color-singlet vector mesons ($\rho_T$ and $\omega_T$) and pseudo-scalar mesons 
($\pi^0_T$ and $\pi^\pm_T$ also dubbed technipions). 
Cross-sections and branching fractions depend in particular on the $\rho_T$ and $\omega_T$ masses, on their mass difference 
with the technipions, and on two mass parameters: $M_A$ for the axial-vector and $M_V$ for the vector couplings which
are set here to $M_V=M_A=500$~GeV.
\subsection{Event selection and analysis}
The analysis looks for production of $\rho_T$ subsequently decaying as
$\rho_T^\pm \rightarrow W^\pm(e\nu) \pi_T^0(b\bar b)$ or
$\rho_T^0 \rightarrow W^\mp(e\nu) \pi_T^\pm(c\bar b/b\bar c)$. The first step is to 
select $W(e\nu)+$~heavy-flavor jets events. 
One requires exactly one electron with transverse momentum $p_T>20$~GeV and pseudo-rapidity $|\eta|<1.1$, 
missing transverse energy $\met >20$~GeV and transverse mass $M_T>30$~GeV, two jets with $p_T>20$~GeV and $|\eta|<2.5$
with at least one of them $b$-tagged.
The SM backgrounds, namely $t\bar t$ production and 
$W/Z$ produced in association with heavy flavor jets are estimated using Monte Carlo simulations (MC). Multi-jet 
production with a jet mis-identified as an electron, and $W$ + light flavor 
production with light jets mistagged as heavy jets comprise the instrumental background which is estimated from data.
At that level there is a good agreement between the 115.1 events estimated for the sum of the backgrounds 
and the 117 found in data. Two startegies are then used to try to extract the signal, one is cut based (CB) and the 
other uses a neural network (NN).

The CB analysis uses kinematic and topological quantities to discriminate against $t\bar t$ and $W+jets$ production.
The invariant mass of the dijet system $M_{jj}$ is used to get indication of a $\pi_T$ narrow resonance. The invariant 
mass of the $W+dijet$ system $M_{Wjj}$ is used to look for a $\rho_T$ narrow resonance. A mass dependant optimization 
is performed on signal significance. For example, for $M_{\pi_T}=110$~GeV and 
$M_{\rho_T}=210$~GeV, 12 events are seen in data for $12.7\pm 0.9$ expected from backgrounds
and $10.3\pm 1.0$ from signal.
%\begin{figure}[h]
%\centerline{\psfig{file=H20F14.ps,width=2.0in}}
%\caption{Distribution of the $Wjj$-system invariant mass}
%\label{fig1}
%\end{figure}

\begin{figure}[h]
\centerline{\psfig{file=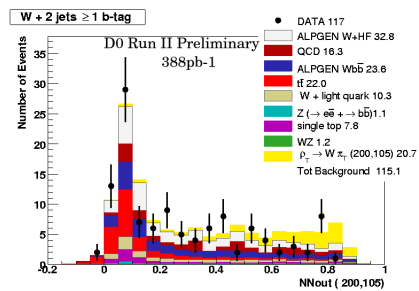,width=2.2in}}
\caption{Distribution of the output of the Neural Network in the $\rho_T\rightarrow W\pi_T$ analysis}
\label{fig2}
\end{figure}

The NN analysis uses a two stage NN using 8 kinematic and topological variables to discriminate signal from 
$t\bar t$ and $W+jets$ production. A mass dependant optimization 
of the NN output cut is performed w.r.t. signal significance. 
An example of NN output is shown in Fig.~\ref{fig2} for $M_{\pi_T}=105$~GeV and 
$M_{\rho_T}=200$~GeV. 

\subsection{Results}
\begin{figure}[h]
\centerline{\psfig{file=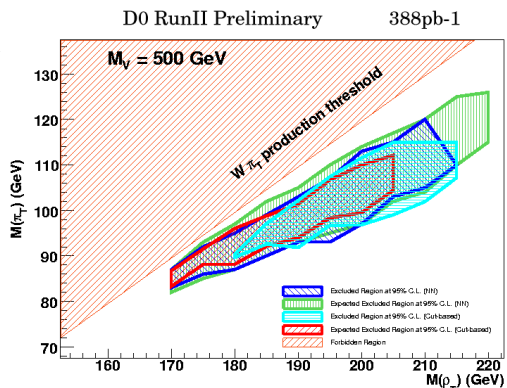,width=2.9in}}
\caption{Observed and expected 95\% C.L. exclusion contours in the $M(\pi_T) vs. M(\rho_T)$ for the Cut and 
Neural Network based analyses.
Regions excluded lie inside the corresponding contours}
\label{fig3}
\end{figure}

Since no excess of events is found in either analysis, 
limits are computed using Bayesian statistics (CB) and 
a 2-D maximum likelihood using ($M_{Wjj}$,$M_{jj}$) correlations (NN). Figure~\ref{fig3} shows the observed and 
expected 95\% C.L. exclusion contours for each analysis.

This is the first measurement done in the TCSM2 framework. No evidence of production of techni-particules was 
found in 388~pb$^{-1}$ of data. Including the $\mu$ channel and taking advantage of the increasing available luminosity
should allow for an even larger TC parameter space exploration.

\section{Search for Leptoquarks}
%\subsection{Introduction}
Leptoquarks (LQ), exotic scalar or vector particles carrying the quantum numbers of a quark-lepton system, are predicted 
by some extensions to the SM which try to relate the apparent symmetry of 
the quark and lepton sectors. To avoid unacceptably large FCNC processes, the LQ's would come in three generations, each 
one coupling to a specific quark/lepton generation. They are expected to decay with a branching fraction $\beta$ 
into a quark and a charged lepton 
and $(1-\beta)$ into a quark and a neutrino. At the Tevatron, if sufficiently light, they can be 
pair produced with a cross-section independant of the unknown LQ-quark-lepton coupling.
First~\cite{1stGene} and second~\cite{2ndGene} generation LQ searches in Tevatron Run II data have already been 
published by D0.
  
When both LQ's decay into a quark and a neutrino, the final state consists of 2 acoplanar jets and \met . A first analysis 
is presented looking for LQ's of any generation in that final state. The second looks specifically for the 3rd generation 
taking advantage of the fact that in this case the jets come from the hadronization of $b$ quarks and thus can be 
tagged as such. Only scalar LQ's are taken into account here since they have a smaller and 
less model dependant production cross-section.

Both analyses use 310~pb$^{-1}$ of Tevatron Run II data recorded by D0, resulting in about 14 million events 
collected with a specific jets+\met\ trigger.

Backgrounds from SM processes ($W/Z$ production associated with jets, diboson production, single- and pair-top 
production) are determined from MC simulations. The instrumental (also dubbed 'QCD') background in multijet production 
is estimated from data.

\subsection{Leptoquarks in the acoplanar jet topology analysis}
\begin{figure}[h]
\centerline{\psfig{file=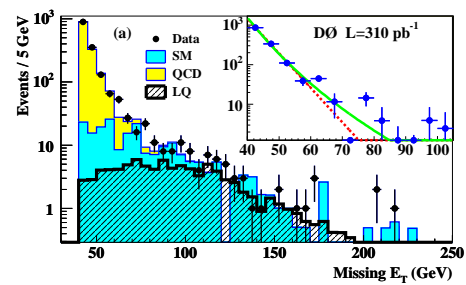,width=2.5in}}
\caption{Distribution of \met\ , after all cuts but \met\ applied, and showing the good agreement between data 
and SM background at high \met\ and how the instrumental background is estimated.}
\label{fig4}
\end{figure}

The selection criteria consist of: a rejection of events with obvious calorimeter noise, requirements on jet properties 
(acoplanarity $>165^o$ between the 2 leading jets, $|\eta|<1.5$ since the signal is central, 
$p_{Tj1}>60$~GeV and $p_{Tj2}>50$~GeV, energy fraction in the electromagnetic
calorimeter $< 0.95$, to reject jets likely due to photons and electrons, 
charged particle fraction $> 0.05$, to avoid fake jets 
and wrong interaction vertices), a rejection of events with isolated electrons or muons with $p_T>10$~GeV, or isolated 
track with $p_T>5$~GeV. To further reduce the backgrounds, exactly 2 jets are required.
A \met\ cut and cuts on angular correlations between jets and the direction 
of \met\ are optimized as a function of the LQ mass ($M_{LQ}$) and used to suppress both SM and instrumental backgrounds.
The remaining instrumental background is estimated from extrapolations from fits to the \met\ distribution in 
the $[40,60]$~GeV interval as shown in Fig.~\ref{fig4}.

As an example, after cuts optimized for $M_{LQ}=140$~GeV, 86 events are observed 
for $72.9^{+10.1}_{-9.7}(stat.)^{+10.6}_{-12.1}(syst.)$ expected from SM backgrounds, 
$2.3\pm1.2$ estimated instrumental background, and $51.8\pm 1.8(stat.)^{+5.6}_{-4.6}(syst.)$ signal events.

\begin{figure}[h]
\centerline{\psfig{file=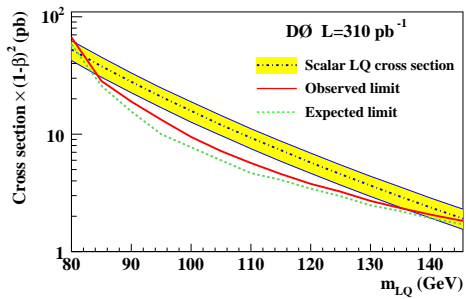,width=2.5in}}
\caption{Observed (solid) and expected (dash) 95\% C.L. excluded cross-section $\times (1-\beta)^2$ as a function of 
$M_{LQ}$, compared to the theoretical cross section for $\beta=0$ (dash-dot) also showing the effect of systematic 
uncertainties (colored band).}
\label{fig5}
\end{figure}

Since no excess is seen, the observed 95\% C.L. excluded cross-section as a function of $M_{LQ}$ is compared 
to the theoretical cross-section reduced by the renormalization scale and the PDF uncertainties summed in quadrature 
as shown in Fig.~\ref{fig5}.

Their intersection allows to set a limit $M_{LQ}>136$~GeV, the most stringent limit to date for 1st and 2nd generation LQ's
for $\beta=0$. This result is now published~\cite{allGene}.

\subsection{Third generation Leptoquarks (LQ$_3$) analysis}

\begin{figure}[h]
\centerline{\psfig{file=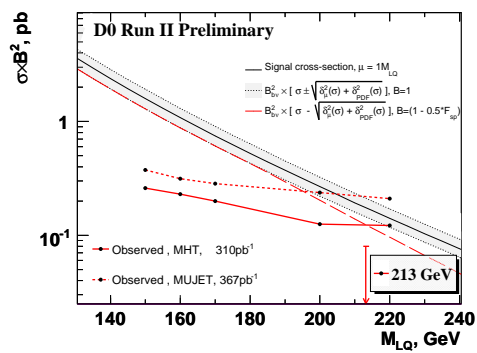,width=2.5in}}
\caption{Observed 95\% C.L. limit on $\sigma B^2(LQ_3\rightarrow b\nu)$ (points and solid line) as a function of $M_{LQ}$
compared to the theoretical predictions (solid line) including systematic uncertainties (colored band),
with and without taking $LQ_3\rightarrow t\tau$ into account.}
\label{fig8}
\end{figure}

In addition to applying selection criteria very similar to the generic LQ search described above, the 2 jets are 
required to be $b$-tagged. After all cuts, for $M_{LQ}=200$~GeV, 1 event is observed for $3.47\pm 0.24(stat.)$ 
expected from SM background and $8.8\pm 0.2(stat.)$ expected from signal. Since no excess of events is seen and since 
the contribution from the instrumental background is estimated to be very small, it is conservatively neglected 
in setting limits for the production of 3rd generation LQ's. Including systematic uncertainties, the observed 
95\% C.L. excluded cross-section can be compared to theoretical predictions (Fig.~\ref{fig8}), where the
fact that when
$M_{LQ}>m_t+m_\tau$, LQ$_3$ could decay in $t\tau$ in addition to $b\nu$, is also taken into account. The result is
a $M_{LQ}$ limit of 213~GeV when the $t\tau$ decay is open and 219~GeV otherwise.
%\begin{figure}[h]
%\centerline{\psfig{file=N4701a.ps,width=2.2in}}
%\caption{}
%\label{fig6}
%\end{figure}

%\begin{figure}[h]
%\centerline{\psfig{file=N4702a.ps,width=2.2in}}
%\caption{}
%\label{fig7}
%\end{figure}

\end{document}